\documentstyle[12pt]{article}
\begin{document}
\newcommand{\lb}{\label}
\newcommand{\ga}{\gamma}
\newcommand{\be}{\begin{equation}}
\newcommand{\ee}{\end{equation}}
\newcommand{\beqa}{\begin{eqnarray}}
\newcommand{\eeqa}{\end{eqnarray}}
\newcommand{\fr}{\frac}
\newcommand{\D}{\delta}
\newcommand{\ebz}{\bar{\eta}_0}
\newcommand{\ebp}{\bar{\eta}_+}
\newcommand{\ebm}{\bar{\eta}_-}
\newcommand{\sig}{\sigma}
\newcommand{\del}{\partial}
\newcommand{\wv}{\wedge}
\newcommand{\al}{\alpha}
\newcommand{\la}{\lambda}
\newcommand{\La}{\Lambda}
\newcommand{\ep}{\epsilon}
\newcommand{\pr}{\prime}
\newcommand{\ti}{\tilde}
\newcommand{\om}{\omega}
\newcommand{\Omo}{\Omega}
\newcommand{\bi}{\bibitem}
\newcommand{\ba}{\begin{array}}
\newcommand{\ea}{\end{array}}

\rightline{IC/93/5}

\rightline{January, 1993}

\vspace{2cm}

\begin{center}

{\Large
BFV--BRST Analysis of the Classical and Quantum

\vspace{.5cm}

$q$-deformations of the $sl(2)$ Algebra}

\vspace{2cm}

{\normalsize
\"{O}mer F. DAYI}

{\small \it
University of Istanbul,
Faculty of Science,  \\
Department of Physics,
Vezneciler, 34459 Istanbul, Turkey, \\
and \\
International Centre for Theoretical Physics,\\
 P.O.Box 586, 34100-Trieste, Italy}

\vspace{1.5cm}

{\bf Abstract}

\vspace{.5cm}

\end{center}

BFV--BRST charge for $q$-deformed algebras
is not unique. Different constructions of it in the
classical as well as in the quantum phase space for the $q$-deformed
algebra $sl_q(2)$ are discussed. Moreover, deformation of
the phase space without deforming the generators of
$sl(2)$ is considered. $\hbar$-$q$-deformation of the
phase space is shown
to yield the Witten's second deformation.  To study the
BFV--BRST cohomology problem when
both the quantum phase space and the group
are deformed, a two parameter
deformation of $sl(2)$ is proposed, and  its BFV-BRST charge
is given.

\pagebreak

\section{Introduction}

Gauge symmetries of a lagrangian manifest themselves as
first-class constraints in the hamiltonian framework. These constraints
can be emploied in constructing a fermionic, nilpotent operator, known
as  Batalin--Fradkin--Vilkovisky--Becchi--Rouet--Stora--Tyutin (BFV--BRST)
charge \cite{BFV}, after quantizing the related phase space and introducing
ghost variables (fields). Although ghost variables are
an artifact of quantization procedure, they can be incorporated
into classical mechanics by endowing classical phase space
with the  generalized Poisson brackets. Hence it appears that one can
establish a BFV--BRST charge and study its cohomology
either in quantum or in classical framework \cite{HT}. In
constrained Hamiltonian systems BFV--BRST cohomology classes are
equated with physical
states. Therefore one can utilize BFV--BRST charge to define a gauge theory.

First class constraints are in involution on the constraint surface,
and in Yang--Mills theories
they constitute a Lie algebra after quantization (in
the absence of anomalies). This suggests to study the
$q$-deformed algebras \cite{Dri}-\cite{rus} in a similar manner, to
elucidate their structure and to extract some clues useful in formulating
$q$-deformed gauge theories.
Obviously, quantization ($\hbar$-deformation) and
$q$-deformations are distinct \cite{AV},\cite{FLS}.
Hence, BFV--BRST analysis of $q$-deformations can be accomplished
either in classical or quantum framework.

There are some attempts to formulate a $q$-deformed gauge
theory \cite{AV},\cite{qYM}, but a complete understanding is lacking.
The study of the
BFV--BRST  structure of $q$-deformed systems in terms of quantum as
well as classical mechanics can facilitate      construction of
the desired gauge theories.

When we deal with the Lie algebras or with the usual constrained
systems, construction of BFV--BRST charge is unique
up to canonical transformations. For
$q$-deformed algebras it depends on the differential
calculus adopted over the group or on the behaviour of constraints.
In Refs.\cite{KMGK}-\cite{JWYZ} two different possibilities are considered.

Properties of the $\hbar$-$q$-deformed algebras are well established,
but  to investigate the classical case (i.e. only $q$-deformation)
there are some
different methods \cite{AV},\cite{FLS},\cite{res},\cite{Suz}.
A powerful way of defining quantum mechanics
as $\hbar$-deformation of the classical one is to utilize
the Moyal $\star$  product\cite{Boy}.
Then a $\star$ product can  be used to acquire
$q$-deformations of classical phase space.
$\star$  product also establishes
the distinction of $\hbar$- and $q$-deformations obviously. Moreover,
it is useful in introducing multiparameter deformations. In fact,
in Section 2 this construction of ``$q$-classical mechanics"
is used to obtain a BFV--BRST charge  for $sl(2)$ generators. Moreover,
there we deal with the phase space endowed with the usual Poisson
brackets but a $q$-deformed ``classical $sl(2)$" algebra.

In Section 3 we perform
$\hbar$-deformation of the cases studied in Section 2.
It is shown that a realization of BFV--BRST charge
for $U_q(SU(2))$ yields
a formulation of BRST cohomology similar to the usual case
\cite{GBY},\cite{Hol}.
When the phase space is $\hbar$-$q$-deformed the usual
$sl(2)$ generators lead to the Witten's second
deformation\cite{Wit}.
In Section 4 a two parameter deformation of $sl(2)$
is introduced
to study the BFV-BRST cohomology problem when both the phase space and
the algebra are deformed, and the related BFV-BRST charge is given.

\section{Classical BFV--BRST Charges}

We deal with a 1-d system (the usual time coordinate),
and ${\bf R}^2$ phase space.
In terms of the phase space variables $(p,x)$,
satisfying the usual
Poisson brackets
\[
\{p,x\}=1,
\]
the ``classical $sl(2)"$ algebra
\be
\lb{sl2}
\{H^0,X^0_{\pm}\}=\pm 2X^0_{\pm},\   \{X^0_+,X^0_-\}=H^0,
\ee
can be realized if the generators are taken to be
\be
\lb{gsl2}
H^0=2px,\  X^0_+=-\sqrt{2} x,\   X^0_-=\fr{1}{\sqrt{2}} p^2x.
\ee
We consider ``$q$-classical systems" defined as \\
{\bf 1)}Poisson brackets are standard, nevertheless
the ``classical $q$-deformed algebra $sl_q(2)$"
is functionally realized  in $C^\infty ({\bf R}^2)$.\\
{\bf 2)}The phase space is endowed with  $q$-deformed Poisson
brackets, but the generators are as in (\ref{gsl2}).

{\bf 1)} In the phase space endowed with the usual
Poisson brackets a functional realization of
the ``classical $sl_q(2)$"
\be
\lb{qusl2}
\{H,X_{\pm}\}=\pm 2X_{\pm},\   \{X_+,X_-\}=\fr{q^H-q^{-H}}{q-q^{-1}}
\equiv [H]_q,
\ee
can be achieved in terms of\cite{FLS}
\be
\lb{qx-}
H=2px,\  X_+=-\fr{1}{2}x, \   X_-=\fr{1+\cosh (2\al px)}{x\al\  \sinh \al },
\ee
where $q\equiv e^{\al}$.

Let us introduce some ghost variables by enlarging the classical phase
space  endowed it with a generalized Poisson bracket structure,
to write a BFV--BRST charge. How many ghost fields are needed?
In \cite{JWYZ} three ghost variables (and their momenta) are used
demanding that the related $q$-$\hbar$-deformed BFV--BRST charge
would be  a
polynomial in
$q^H$.  Although this is quite plausible
(in comultiplication of $sl_q(2)$, $q^H$ appears),
it is not the unique choice: the form (or the number) of
constraints will dictate the number of ghost variables.
In Ref.\cite{JWYZ}   it is assumed that there are three constraints
behaving as $X_\pm$, and $[H]_q$.
But a priori one does not know the structure of the constraints. There
may be different choices: in Ref.\cite{AV} a candidate for
a $q$-deformed gauge theory is shown to possess infinite gauge field
components (hence infinite constraints) depending on the representation
of the universal covering algebra. Another deformation of the BRST
algebra is given in Ref.\cite{Wat} where the number of the ghost
fields is 4. Let us deal with the cases where there are three
ghost variables, but the assumed constraint structures are
different from Ref.\cite{JWYZ}.

After choosing  three ghost variables and their momenta, we should
also define generalized
Poisson brackets of them. This depends on the conditions
which we require that  BFV--BRST charge satisfies.

To assume that the constraints behave as
$X_\pm$, and $H$, seems to be the simpliest choice. By using
\beqa
\fr{e^{\al H}-e^{-\al H}}{e^\al -e^{-\al }} &
=  &  H   \fr{\al }{e^\al -e^{-\al }} \sum_{k=0}^\infty
\fr{(\al H)^{2k}}{(2k+1)!}  \nonumber \\
 & =  & H   \fr{\al }{e^\al -e^{-\al }}
 \prod_{k=1}^\infty (1+ \fr{\al^2 H^2}{k^2 \pi^2})  \\
  &  \equiv  & H f(H,q),    \nonumber
\eeqa
and introducing the fermionic (ghost) variables $(c^i,\ \pi_i),\
i,j=0,+,-,$
which satisfy the usual generalized Poisson
brackets
\be
\lb{gh}
\{ \pi_i, c^j \} =\D^j_i,\
\{ \pi_i,\pi_j \}=0,\
\{ c^i,c^j \}=0,
\ee
one can write the classical BFV--BRST charge as
\be
\lb{b1}
\Omo_1
=c^+ X_+ +c^-X_-+\fr{1}{\sqrt{2}}
c^0H -\sqrt{2} f(q,H)c^+c^-\pi_0
+\sqrt{2}c^+c^0\pi_+ -\sqrt{2}c^-c^0\pi_-.
\ee
The generalized Poisson brackets are
\[
\{ f,g \}=
\fr{\del f}{\del p}\fr{\del g}{\del x}
-\fr{\del f}{\del x}\fr{\del g}{\del p}
+\fr{\del f}{\del \pi_i}\fr{\del g}{\del c^i}
+\fr{\del f}{\del c^i}\fr{\del g}{\del \pi_i} .
\]

One can easily observe that $\Omo_1$ satisfies the classical nilpotency
relation \[
 \{ \Omo_1 ,\Omo_1 \}  =0.
\]

We suppose that
the generalized Poisson brackets of the ghosts
are non-deformed
due to the fact that we did not deform the original phase space.
But the ghost variables are associated with the gauge (group)
generators, so that deforming their Poisson brackets, even if
the original phase space is not deformed, is not ruled out.

Another possibility is to suppose that the constraints behave as
$X_\pm $, and $\left[ \fr{H}{2} \right]_q$.
The choice where $\left[ \fr{H}{2} \right]_q$ is replaced by
$\left[ {H} \right]_q$ seems more natural because in the coproduct
of $sl_q(2)$,  $q^H$ appears. Nevertheless, in the following
section we show that our choice possesses  more similarities
with the usual BFV--BRST analysis.
The related classical
BFV--BRST charge
satisfying
\[
{ \{ \Omo_2,\Omo_2 \} } =0,
\]
is given by
\beqa
\Omo_2 & = & X_+c^++X_-c^- +(q+q^{-1})^{1/2}{\left[ \fr{H}{2} \right]_q}
c^0  -
(q+q^{-1})^{-1/2} \left( \fr{q^H-q^{-H}}{q^{H/2}-q^{-H/2}} \right)
\pi_0c^+c^-
\nonumber \\
 & & +\fr{ln\ q}{ (q-q^{-1})} (q+q^{-1})^{1/2}(q^{H/2}+q^{-H/2})
(\pi_+c^+c^0 -\pi_-c^-c^0) \nonumber \\
  & & - ln^2 q\  (q+q^{-1})^{1/2}
{\left[ \fr{H}{2} \right]_q}
 \pi_+\pi_-c^-c^+c^0.            \lb{2}
\eeqa

\vspace{1.5cm}

{\bf 2)}  As announced before a $\star$ product approach is prefered
to $q$-deform the phase space (we  follow  Ref.\cite{Suz}).

Attach a two dimensional internal space parametrized by
$\xi$, and $\rho$, to each point of the phase space by defining
\be
x_\xi =xe^{i\ga \xi},\
p_\rho =pe^{i\ga \rho}.
\ee
Then define a $\star$ product of any functions $f$ and $g$ as
\be
f\star_\ga g\equiv
\sum_{n=0}^\infty \fr{(-\ga/2)^n}{n!}
\sum_{k=0}^n
\left(
\ba{c}
n \\
k
\ea
\right) (-1)^k (\del^{n-k}_\xi \del_\rho^k\  f)\
(\del_\rho^{n-k} \del_\xi^k\ g ).
\ee
This $\star$ product is associative and can be used to define the
$q$-deformed Poisson brackets
\beqa
\{f,g\}_0^\ga & \equiv & -2 \fr{f\star_\ga g -g\star_\ga f}{xp\ ln\ q}
\nonumber \\
& = &
\fr{-2}{xp\ ln\ q}\sum_{n=0}^\infty \fr{ (-\ga /2 )^{2n+1} }{ (2n+1)! }
\sum_{k=0}^{2n+1}
\left(
\ba{c}
2n+1 \\
k
\ea
\right) (-1)^k  \nonumber \\
 &  &  (\del^{2n+1-k}_\xi \del_\rho^k\  f)
 (\del_\rho^{2n+1-k} \del_\xi^k\ g ),
\eeqa
where $q\equiv \exp (\ga^3)$.
Let us deal with the functional realization of classical $sl(2)$,
given in (\ref{qx-}) by replacing
$x\rightarrow x_\xi,\ p\rightarrow p_\rho$:
\[
H^\ga =2p_\rho \star_\ga x_\xi ,\
X_+^\ga =-\sqrt{2} x_\xi ,\
X_-^\ga =\fr{1}{\sqrt{2}} p_\rho \star p_\rho \star x_\xi .
\]
These satisfy the following $q$-deformed Poisson brackets in the
limit $\rho ,\xi \rightarrow 0$,
\[
\{H,X_{\pm}\}_0^\ga =\pm 2a(q) X_{\pm},   \{X_+,X_-\}^\ga_0= (a(q)/2)
(q^{1/2}+q^{-1/2} )H,
\]
where
\[
a(q)=\fr{1- q^{-1}}{ln\  q}.
\]
It is a Lie algebra in terms of the new brackets, thus
we obviously need three ghost variables and their momenta
for the BFV--BRST analysis.
The generalized Poisson brackets of the fermionic
ghost variables should be deformed or not? In the
$q$-$\hbar$-deformed case there is somehow a natural
answer to this, but at the
classical level it is completely arbitrary. Hence we suppose that
they satisfy the usual conditions
\be
\lb{cc}
c^ic^j =-c^jc^i,\  \pi_i \pi_j =-\pi_j\pi_i,\  i\neq j,
\ee
and
\be
\lb{cp}
\{c^i,\pi_j\}^\ga=\D^i_j.
\ee
Then, the generalized $q$-deformed Poisson brackets are
\be
\lb{qgp}
\{f,g\}^\ga \equiv -2 \fr{f\star_\ga g - (-)^{\ep (f) \ep (g)}
g\star_\ga f}{xp\ ln\ q}
+\fr{\del_lf}{\del \pi_i}\fr{\del_r g}{\del c^i}-
(-)^{\ep (f)\ep (g)}\fr{\del_lg}{\del \pi_i}\fr{\del_r f}{\del c^i},
\ee
where $\ep (f)$ indicates the ghost number :
\[
\ep (c^i) =-\ep (\pi_i)= 1,\     \ep (fg) =\ep (f) +\ep (g).
\]
Hence we write the BFV--BRST charge  as
\beqa
\Omo_3 & = & Hc^0+X_+c^++X_-c^-- a(q)\pi_+c^0c^++a(q)\pi_-c^0c^-
\nonumber \\
& &  -(a(q)/2) (q^{1/2}+q^{-1/2})
\pi_0c^+c^-, \lb{b3}
\eeqa
which satisfies
\[
\{\Omo_3,\Omo_3\}^\ga =0.
\]

In the next section we show that in $\hbar$-$q$-deformed case,
it is natural to keep (\ref{cp}) but deform (\ref{cc}). If we
adopt a similar deformation in the classical case a BFV--BRST charge
which possesses terms
linear in the generators as in (\ref{b3}), will not exist.

\section{Quantization}

When we deal with the non-deformed phase space,
there is no difference between introducing the
$\hbar$-deformation in terms of
the Moyal brackets or the canonical quantization as far as
the purposes of this section  are considered.
If we drop $\star$  in the former formulation, both of the them
will yield the following fundamental commutators
\be
\lb{fc}
[p,x]=-i \hbar.
\ee
Of course, when (\ref{fc}) is considered as Moyal brackets $p$ and $x$
are classical variables, but they are operators in terms of
the canonical quantization.

After an appropriate rescaling of the generators, (\ref{sl2})
becomes the usual $sl(2)$ algebra and (\ref{qusl2}) reads \cite{FLS}
\be
\lb{qhusl2}
{[}H,X_{\pm}]=\pm 2X_{\pm},  \    [X_+,X_- ]= [H]_q.
\ee

The ghost fields, then, satisfy
\[
[\pi_i,c^j]=-i\hbar\D^j_i,
\]
where $[f,g]=fg-(-)^{\ep (f)\ep (g)}gf$.
For simplicity we rescale the phase space variables such that
\[
[p,x]=1,\    [\pi_i,c^j]=\D_i^j.
\]
Under the $\hbar$-deformation
$\Omo_1 \rightarrow Q_1$ which is in the same form but satisfying
\be
\lb{q1}
Q^2_1=0.
\ee

The BFV--BRST charge for $U_q(SU(2))$, (\ref{qhusl2}),
$Q_2$, satisfying $Q_2^2=0$, when the constraints
are supposed to behave like $X_\pm$, and $\left[ \fr{H}{2}\right]$, is not any
more given similar to (\ref{2}), but
\beqa
Q_2  & = & X_+c^++X_-c^-
+(q+q^{-1})^{1/2}{\left[ \fr{H}{2} \right]_q} c^0  -
(q+q^{-1})^{-1/2} \fr{q^H-q^{-H}}{q^{H/2}
-q^{-H/2}} \pi_0c^+c^-  \nonumber \\
 &  & +\fr{(q+q^{-1})^{1/2}}{q^{1/2} -q^{-1/2}}
 \{ (q^{(H+1)/2}+q^{-(H+1)/2})\pi_+c^+c^0
 -(q^{(H-1)/2}+q^{(-H+1)/2})\pi_-c^-c^0)\}
\nonumber \\
 & &  - \fr{(q+q^{-1})^{1/2}}{(q^{1/2}+q^{-1/2})^2}
 (q-q^{-1})^2 {\left[ \fr{H}{2} \right]_q}
 \pi_+\pi_-c^-c^+c^0. \lb{q2}
\eeqa

To obtain the physical states or the solution of the BRST cohomology,
let us consider the space of the states
\[
\Psi (c) =\sum_{l=0}^3  \fr{1}{l!} c^{i_1} \cdot c^{i_l}
\Psi_{i_1\cdot i_l}^{(l)}.
\]
The $\Psi^{(l)}$ coefficients are some
complex functions on the space where the
constraints or the generators act.
Action of $\pi_i$  on the states is
\[
(\pi_i \Psi )^{(l)}_{i_1\cdot i_l}=\Psi_{ii_1i_l}^{(l+1)};\ l=0,1,2.
\]
When one deals with a Lie algebra the coefficients
$\Psi_{i_1\cdot i_l}^{(l)},$ can be considered as $l$-forms on
the algebra, and the indices are raised or lowered  by
the Cartan metric of the algebra. Thus one can introduce the
scalar product\cite{Hol}
\be
\lb{pd}
(\Phi ,\Psi )=\sum_{l=0}^3 \fr{1}{l!}\Phi^{\dagger (l)i_1\cdot i_l}
\Psi^{(l)}_{i_1\cdot i_l},
\ee
which is positive definite. With respect to this product
\be
\lb{dag}
c^{i\dagger }=\pi_i.
\ee

$Q^\dagger $ obtained from the BFV-BRST charge
$Q$ of the Lie group is also  nilpotent. When we deal with
$SU(2)$ and demand that $[Q,Q^\dagger ]$ is a generalization
of the quadratic Casimir of $SU(2)$, in the basis we adopted
the scalar product should also yield
\be
\lb{her}
X_\pm^\dagger =X_\mp,\  H^\dagger =H.
\ee

In the case where we assume that the constraints behave
like $X_\pm$, $H$, the conjugation defined by
(\ref{dag})-(\ref{her}) leads to $Q_1^\dagger$ which is nilpotent.
Unfortunately, when the constraints are supposed to behave like
$X_\pm$, $\left[ \fr{H}{2}\right]$, $Q_2^\dagger$ obtained from
(\ref{q2}) is not nilpotent.This is due to the fact that
in the former case BFV-BRST charge is insensible to the
ordering of ghost variables, but in the latter a change in the
ordering of ghost variables would create some terms which spoil the
nilpotency condition.

To overcome this difficulty let us introduce the following positive
definite scalar product
\be
\lb{cpd}
(\Phi^*,\Psi )=\sum_{l=0}^3 \fr{1}{l!}\Phi^{*(l)}_{i_l \cdot i_1}
g^{i_1 j_1} \cdot g^{i_l j_l}\Psi^{(l)}_{j_1 \cdot j_l},
\ee
where $g^{00}=g^{+-}=g^{-+}=1$, and   $\Phi^{*(l)}_{i_l \cdot i_1}$
is the complex conjugate of $\Phi^{(l)}_{i_l \cdot i_1}$. With respect
to this product the conjugate of the generators and the ghost variables
are
\be
\lb{conj}
X_\pm^*=X_\pm ,\   H^*=H,\    c^{0*}=\pi_0,\ c^{+*}=\pi_-,\
c^{-*}=\pi_+,
\ee
where $(f^*)^*=f$. Conjugation of $Q_1$
yields the following co-BFV-BRST charge
\be
\lb{cq1}
Q^*_1=\pi_- X_+ +\pi_+X_-+\fr{1}{\sqrt{2}}
\pi_0H -\sqrt{2}f(q,H)\pi_-\pi_+c^0
+\sqrt{2}\pi_-\pi_0 c^--\sqrt{2}\pi_+\pi_0 c^+,
\ee
which is nilpotent and $[Q_1,Q^*_1]$ is a generalization of the
quadratic Casimir of $SU(2)$. This justifies the choice of the
normalization factors of the  terms linear in the ghost variables.

In terms of the conjugation given in (\ref{conj}), the
co-BFV--BRST charge derived from $Q_2$ is
\beqa
Q^*_2  & = &
X_+\pi_- +X_-\pi_++(q+q^{-1})^{1/2}{\left[ \fr{H}{2} \right]_q} \pi_0  -
(q+q^{-1})^{-1/2} \fr{q^H-q^{-H}}{q^{H/2}-q^{-H/2}}
c^0 \pi_-\pi_+ \nonumber \\
 &  & +\fr{(q+q^{-1})^{1/2}}{q^{1/2} -q^{-1/2}}
 \{ (q^{(H+1)/2}+q^{-(H+1)/2})c^-\pi_-\pi_0
 -(q^{(H-1)/2}+q^{(-H+1)/2}) c^+\pi_+\pi_0)\}
\nonumber \\
 & &  - \fr{(q+q^{-1})^{1/2}}{(q^{1/2}+q^{-1/2})^2}
 (q-q^{-1})^2 {\left[ \fr{H}{2} \right]_q}
c^- c^+\pi^+\pi^-\pi^0. \lb{cq2}
\eeqa
This charge as in the usual case,  can be used to define
\be
\lb{ham}
\{ Q_2,Q^*_2\}|_{\pi=c=0}=C_q,
\ee
where $C_q$ is the quadratic Casimir of $U_q(SU(2))$\cite{Cas},\cite{CZ}:
\[
C_q=X_-X_++
\left[ \fr{H}{2}\right]_q
\left[ \fr{H+2}{2}\right]_q=
\fr{1}{2} \left( X_-X_++X_+X_- + (q+q^{-1})
\left[ \fr{H}{2}\right]_q
\left[ \fr{H}{2}\right]_q
\right).
\]

Hence by using the positive definite scalar product (\ref{cpd}), the physical
states can be identified with the states $\omega$ satisfying
\[
(Q+Q^*)^2\omega =0,\   Q\omega =0,    Q^* \omega=0,
\]
where $Q$ and $Q^*$ are given either by (\ref{q1}) and (\ref{cq1})
or by (\ref{q2}) and (\ref{cq2}). At zero ghost number
the cohomology classes given by (\ref{q2}) include the ones
found in Ref.\cite{JWYZ}, and
the states $\omega_2$ satisfying $Q_2\omega_2=0$,
contain the singlets of $U_q(SU(2))$.
Although at zero ghost number $Q_1\omega_1=0$
yields the states $\omega_1$, which are singlets of  $SU(2)$, by including
ghost number one sector the other states of $U_q(SU(2))$ can be
obtained.

 If we $\hbar$-deform the phase space after the $q$-deformation we get
 \cite{Suz}
\be
\lb{hqd}
x\star_{\ga\hbar}p-qp\star_{\ga\hbar}x=-i\hbar q^{1/2} .
\ee
The $\star_{\ga\hbar}$ product is defined as
\beqa
f(x,p)\star_{\ga\hbar}g(x,p) & = &
f(x_\xi,p_\rho )
\sum_{n=0}^\infty \fr{(-\ga/2)^n}{n!}
\sum_{k=0}^n
\left(
\ba{c}
n \\
k
\ea
\right) (-)^k (\del^{n-k}_\xi \del_\rho^k )
 (\del_\rho^{n-k} \del_\xi^k) \nonumber \\
& & \sum_{m=0}^\infty \fr{(-i\hbar /2)^m}{m!}
\sum_{l=0}^m
\left(
\ba{c}
m \\
l
\ea
\right) (-)^l (\del^{m-l}_x \del_p^l )\
(\del_p^{m-l} \del_x^l)\
  g(x_\xi ,p_\rho ) |_{\xi =\rho =0}, \nonumber
\eeqa
where in the sums the first two derivatives act on the left
and the others on the right hand side.

For our purposes, once the $\hbar$-deformation is achieved
we can forget about the $\star_{\ga\hbar}$ and also set
$\hbar =1$, to obtain
\be
\lb{hqc}
xp-qpx =-iq^{1/2}.
\ee

By keeping the form of the generators as in (\ref{gsl2})
we get
\beqa
HX_--qX_-H & = & -i2q^{1/2}X_- , \lb{w1} \\
HX_+-q^{-1}X_+H & = &  i2q^{-1/2}X_+ , \\
X_+X_--q^2X_-X_+ & = & (-i/2)(q^{1/2} +q^{3/2})H. \lb{w3}
\eeqa
After rescaling as
\[
X_\pm \rightarrow i q^{-1/2}\sqrt{(q^{1/2}+q^{3/2})} ,\
  H \rightarrow i2 H
\]
and setting
\[
q=r^2,
\]
one can see that the relations given in (\ref{w1})-(\ref{w3}) read
\beqa
r^{-1}HX_--rX_-H & = & -X_- , \nonumber \\
rHX_+-r^{-1}X_+H & = &  X_+ , \lb{wq} \\
r^{-2}X_+X_-  -r^2X_-X_+ & = & H, \nonumber
\eeqa
in which we recognize  the Witten's second deformation \cite{Wit}.

When we $q$-$\hbar$-deform the phase space a natural requirment for the
BFV--BRST charge is to demand that the anticommutation of
the terms which are linear in ghost variables and the generators,
with themselves generates the algebra. For the deformed algebra
(\ref{wq}), this condition   leads to
\be
\lb{1f}
c^ic^j=-\nu^{ij}c^jc^i;\ \nu^{ii}=0, \nu^{0+}=r^2,\ \nu^{-0}=r^2,\
\nu^{-+}=r^4. \ee
These relations could also be obtained by demanding that
$c^i$ behave like one forms on $sl_q(2)$ \cite{Wor}.

Although one can also deform the
commutators, it is not necessary. In fact, we deal with the ghost
variables satisfying
\be
\lb{2f}
{[\pi_i,c^i]}=\D^i_i, c^{i2}=\pi^2_i=0.
\ee
Now the associativity leads to
\[
\pi_i \pi_j=-\nu^{ij}\pi_j\pi_i,\   \pi_ic^j=-\nu^{ji}c^j\pi_i.
\]
Hence the BFV--BRST charge which satisfies $Q_3^2=0$, is
\beqa
Q_3= Hc^0+X_+c^++X_-c^--r(\pi_+c^+c^0+\pi_-c^0c^-)
-r^2 \pi_0c^+c^-\nonumber \\
 -(r-r^{-1})\pi_-\pi_+c^0c^+c^-. \lb{bq3}
\eeqa

One can observe that the choice (\ref{1f})-(\ref{2f}) follows if
we require that $Q_3$ behaves like the exterior derivative,
so that $\{Q_3,c^i\}$ coincide with the Cartan-Maurer
structure equations on $sl_q(2)$.

To find solution of the cohomology of $Q_3$ one should define
a state space endowed with a scalar product,
and introduce the co-BFV-BRST charge. A choice is given
in Ref.\cite{KMGK}. The choice should be dictated by the desired
physical content of the gauge theory.
This is still obscure, so that the issue of defining a scalar
product and co-BFV-BRST charge is not discussed here.

\section{Two Parameter Deformation}
Recently, attention is paid to multiparameter deformations of
groups \cite{mpd}. Because of the fact that the requirements are
different, not all of these deformations fulfill the condition of
being an Hopf algebra.

In the procedure which we follow,
obviously, one can $q$-deform the $\hbar$-deformed phase space
as well as the $\hbar$-deformed algebra with different
parameters. In this section $q$-deformation of the quantum
phase space is supposed to be realized as given in Ref.\cite{Jim},
which is known to be equivalent to
the Witten's deformation (\ref{wq}), \cite{CZ}.
If we demand to obtain one of the deformations
at some  special values of the deformation parameters, it is quite natural
to consider the following two parameter deformation of $sl(2)$,

\beqa
\mu^2HX_+-\fr{1}{\mu^2}X_+H & = & (1+\mu^2)X_+ ,\nonumber \\
\mu^2X_-H-\fr{1}{\mu^2}HX_- & = & (1+\mu^2)X_- ,\lb{2da} \\
\fr{1}{\mu}X_+X_--\mu X_-X_+ & = & [H]_q .\nonumber
\eeqa

To keep the resemblance with $\mu=1$ and $q=1$ cases we introduce
the ghost fields satisfying
\be
[\bar{\eta}_i, \eta^i] =\D^i_i,
\ee
\be
\lb{qdf}
\eta^i\eta^j=l_{ij}\eta^j \eta^i,\
\bar{\eta}_i\bar{\eta}_j=l_{ij}\bar{\eta}_j \bar{\eta}_i,\
\eta^i\bar{\eta}_j=l_{ji}\bar{\eta}_j \eta^i,\
\ee
where
\[
l_{0+}=\mu^4,\ l_{-0}=\mu^4,\  l_{-+}=\mu^2,\  l_{ii}=0.
\]

The BFV--BRST charge which leads to the one given in
Ref.\cite{JWYZ} for $\mu =1$ and to the one given
in Ref.\cite{KMGK} for $q=1$, moreover satisfying $Q^2=0$, is
\beqa
Q & = & X_-\eta^- + X_+\eta^+ +
[H]_q\eta^0 -\mu \bar{\eta}_0\eta^+\eta^-
 -F(H)\ebp c^+c^0 \nonumber \\
 & & +G(H) \ebm c^0c^-
   -(G(H) +F(H))\ebm \ebp c^+c^0c^-,    \lb{2ph}
\eeqa
where
\beqa
F(H) & = & \fr{\mu^2}{q-q^{-1}} [\mu^{-2}
(q^H-q^{-H})-\mu^2(q^a-q^{-a})], \nonumber \\
G(H)  & = & \fr{\mu^2}{q-q^{-1}} [\mu^{2}
(q^H-q^{-H})-\mu^{-2}(q^b-q^{-b})], \nonumber \\
a  & = & \mu^{-2}(\mu^{-2}H+\mu^2+1),  \nonumber \\
b & = & \mu^{2}(\mu^{2}H -\mu^2-1). \nonumber
\eeqa

To find  solutions of the $Q$-cohomology, one should
introduce a state space and a scalar product. Obviously, there
are different choices and as it is mentioned above, it is
closely related to the desired physical properties of the system.

\vspace{1cm}

\begin{center}
{\bf \large Acknowledgments}
\end{center}

I thank
Professor Abdus Salam, the International
Atomic Energy Agency and UNESCO for hospitality at the ICTP.

This work is partially supported by the Turkish Scientific and
Technological Research Council (TBTAK).

\end{document}